\RequirePackage{fix-cm}
\documentclass[fleqn,usenatbib]{mnras}

\usepackage{newtxtext,newtxmath}

\usepackage[T1]{fontenc}

\DeclareRobustCommand{\VAN}[3]{#2}
\let\VANthebibliography\thebibliography
\def\thebibliography{\DeclareRobustCommand{\VAN}[3]{##3}\VANthebibliography}

\usepackage{graphicx}	
\usepackage{amsmath}	

\title[A radio counterpart to M82 X$-$1]{An \textit{e}-MERLIN \& EVN radio counterpart to the ultraluminous X-ray source M82 X$-$1}

\author[D.~Williams-Baldwin et al.]{D.~Williams-Baldwin,$^{1}$,\thanks{E-mail: david.williams-7@manchester.ac.uk}
T.~W.~B.~Muxlow$^{1}$,
G.~Lucatelli$^{1}$,
R.~J. Beswick$^{1}$
\\
$^{1}$ Jodrell Bank Centre for Astrophysics, School of Physics and Astronomy, The University of Manchester, Manchester, M13 9PL, UK\\
}

\date{Accepted XXX. Received YYY; in original form ZZZ}

\pubyear{\the\year{}}

\begin{document}
\label{firstpage}
\pagerange{\pageref{firstpage}--\pageref{lastpage}}
\maketitle

\begin{abstract}
Ultra-luminous X-ray sources (ULXs) are X-ray bright (\textit{L}$_{\rm X-ray} >$3$\times$10$^{39}$erg~s$^{-1}$) extra-galactic objects that are powered by either neutron stars, or stellar or intermediate-mass black holes (IMBHs) but few have been detected in the radio waveband. In the nearby galaxy M82, the brightest ULX - M82 X$-$1, is thought to be associated with an IMBH but to date does not have a radio counterpart. We present deep wide-band reprocessed \textit{e}-MERLIN images observed in 2015 May with an r.m.s. sensitivity of 7$\mu$Jy~beam$^{-1}$ and report the discovery of a new radio source with an integrated flux of \textit{S}$_{\rm \nu=4.88\,GHz}$ = 174$\pm$15$\mu$Jy, which is spatially co-incident with the \textit{Chandra} X-ray position of M82 X$-$1. This source is not detected in archival MERLIN/\textit{e}-MERLIN observations in the last three decades. A search for intra-observation variability in the 2015 \textit{e}-MERLIN data was inconclusive, but a comparison with 1.5\,GHz \textit{e}-MERLIN observations taken a week prior yielded no detection. We also detect the source at the same position with milliarcsecond angular resolution in EVN+\textit{e}-MERLIN data from 2021 March at 53$\pm$10$\mu$Jy. The radio source position is ICRF J2000 RA: 09$^{\rm h}$55$^{\rm m}$50$\fs$1172, Dec: +69$^{\circ}$40$\arcmin$46$\farcs$606 ($\pm$1.5\,mas). These radio fluxes are consistent with other radio-detected ULXs on the radio:X-ray plane and points towards a stellar/intermediate-mass black hole. The black hole mass inferred by the `fundamental plane of black hole activity' is 2650~M$_{\odot}$, but this value remains highly uncertain.
\end{abstract}

\begin{keywords}
radio continuum: transients -- radio continuum: galaxies -- galaxies: individual: M82
\end{keywords}



\section{Introduction}

Ultra-luminous X-ray sources (ULXs) are bright (\textit{L}$_{\rm X-ray}$~$>$~3$\times$10$^{39}$~erg~s$^{-1}$) compact extra-galactic off-nuclear X-ray sources \citep[e.g., see][for reviews]{BachettiReview,KaaretReview} commonly found in high star-formation rate galaxies \citep{2009ApJ...703..159S}. This X-ray luminosity corresponds to the Eddington limit of a 20~M$_{\odot}$ black hole \citep{RemillardMcClintock2006}, implying that ULXs could be powered by intermediate-mass black holes (IMBHs) of black hole masses between 100-10$^6$M$_{\odot}$ \citep[e.g.,][]{Mezcua2017IJMPD..2630021M} or super-Eddington accreting stellar-mass black holes. However, the discovery of pulsations in M82 X$-$2 \citep{Bachetti2014} unveiled the first of a population of super-Eddington accreting neutron stars in ULXs. Subsequent searches have revealed other pulsating ULXs and it is possible that many if not all ULXs are powered by neutron stars \citep[e.g.,][]{KingLasota}. Never-the-less, the possibility of finding IMBHs remains a tantalising prospect. One of the most promising IMBH candidates is M82 X$-$1. 

M82 X$-$1 was first discovered as the most luminous X-ray source (\textit{L}$_{\rm X-ray}\sim$10$^{41}$erg s$^{-1}$) in M82 with \textit{Chandra} X-ray observations in 1999 \citep{Kaaret2001}. It is highly variable \citep{Kaaret2009ApJ...692..653K} and due to its distance from the dynamical centre, it is not the actively accreting super-massive black hole at the centre of M82 \citep{Kaaret2001}. Even with super-Eddington accretion, it is difficult to explain M82 X$-$1 as a 
stellar-mass black hole or a neutron star, as the typical conditions of a $\leq$100~M$_{\odot}$ ULX accreting at the Eddington limit can explain luminosities of up to 10$^{40}$erg s$^{-1}$, but above this value an IMBH is required \citep{FengSoria2011NewAR..55..166F}. The X-ray spectral and timing properties obtained using \textit{Chandra} and \textit{XMM-Newton} data showed a thermal state in M82 X$-$1 analogous to those found in Galactic black hole binaries, with a slightly above-Eddington spinning black hole of mass 200$-$800~M$_{\odot}$ preferred \citep{FengKaaret2010}. Futhermore, the discovery of twin-peaked high-frequency quasi-periodic oscillations in the X-ray timing analysis \citep{Pasham2015} argues strongly for a black hole mass of 428$\pm$105~M$_{\odot}$, which agrees with the value obtained (415$\pm$63~M$_{\odot}$) from scaling the relativistic precession model of X-ray binaries \citep[][]{Motta2014}. Multiple black hole mass estimations place a mass of 20$-$1000~M$_{\odot}$ for M82 X$-$1 \citep[see Figure~3 in][and references therein]{Mondal}. All of the evidence above suggests that an IMBH is most likely in this object \citep{KaaretReview}. 

Radio emission has been observed in a handful of ULXs previously. The first ULX with detected radio emission was 2E~1400.2$-$4108 in NGC 5408 \citep{Kaaret2003} with a flux density of \textit{S}$_{\rm \nu=4.8\,GHz}$= 0.26\,mJy which was interpreted as emission from an accreting stellar-mass black hole. ESO 243$-$49 (known as HLX$-$1) is arguably the best IMBH candidate as it has X-ray luminosities of \textit{L}$_{\rm X-ray}\sim$10$^{42}$erg s$^{-1}$ and was found to have flaring radio emission consistent with a discrete ejection event \citep{Webb2012}. \citealt[][]{Mezcua2013MNRAS.436.1546M} detected compact radio emission from the ULX NGC 5457$-$X9 and they suggest it could be an IMBH candidate. Radio emission was detected from a flaring X-ray source in M31 and was attributed to a $\leq$70~M$_{\odot}$ black hole accreting at the Eddington limit \citep{Middleton2013}. Compact and variable radio emission has also been found for Holmberg II X$-$1 \citep{Cseh2014,Cseh2015}, which has a triple structure of radio emission, embedded in a larger radio nebula. Radio nebulae have also been found in other ULXs, for example, IC 342 X$-$1 \citep{Cseh2012}, Holmberg IX X$-$1 \citep{Berghea2020}, NGC 5585 X$-$1 \citep{Soria2020}, NGC 4861 X$-$1 \citep{Gong2023} and NGC 6946 X$-$1 \citep{Beuchert}.

The nearby star-forming galaxy M82 has been studied for decades due to its high star formation rate, enabling investigations of stellar and galaxy evolution. It has multiple compact X-ray sources \citep[e.g., ][]{Kong,Chiang,Iwasawa} including the two aforementioned ULXs M82 X$-$1 and M82 X$-$2. In excess of 50 compact radio sources have been catalogued in M82 as supernova remnants (SNRs), \ion{H}{II} regions \citep[e.g., ][]{Muxlow1994,McDonald2001,Beswick2006,Argo2007,Fenech2008,Fenech2010,Gendre2013} and several `exotic' transient sources using very long baseline interferometry \citep[VLBI, ][]{Kronberg1985,Brunthaler2009,Muxlow41.95,Muxlow2009ATel.2073....1M,Muxlow2010,Brunthaler2010,Joseph2011,Perez-Torres2014,Kimani2016}. In the radio, the first transient discovered, 41.5+59.7 \citep{Kronberg1985} is located 0.8 arcsec from M82 X$-$1, but has since been confirmed to be an unrelated X-ray binary \citep{Xu2015}. Despite efforts to locate a radio counterpart to M82 X$-$1, thus far, none has been found \citep{Kording2005}. 

In this manuscript, we provide the first radio detections of M82 X$-$1 using 50 milliarcsecond (mas) scale images from the \textit{enhanced} Multi-Element Radio-Linked Inteferometer Network \citep[\textit{e}-MERLIN, see ][]{Garrington_2016} and 10 mas scale images from the European VLBI Network (EVN) including \textit{e}-MERLIN here-after `EVN+\textit{e}-MERLIN'. Throughout this work we assume a distance to M82 of 3.2\,Mpc \citep{1964ApJ...140..942B}, equivalent to a linear scale of 15\,pc per arcsec. This paper is structured as follows: In Section 2 we describe our \textit{e}-MERLIN and EVN+\textit{e}-MERLIN data, in Section 3 we show our results and discuss them in the context of the X-ray position of M82 X$-$1, and finally in Section 4, we present our conclusions.

\section{Observations and Data Reduction}

\begin{table*}
    \centering
    \begin{tabular}{c|c|c|c|c|c|c|c|c|c}
         Cent. Obs. date & Central Date & Array & Project &  Band Cent. & Band width & Int. time & Int. Flux & Sensitivity & Note\\
         (yyyy-mm-dd) & (MJD) & & Code & (GHz) & (MHz) & (hrs) & ($\mu$Jy) & $\mu$Jy~beam$^{-1}$ \\
         (1) & (2) & (3) & (4) & (5) & (6) & (7) & (8) & (9) & (10) \\
         \hline
         1992-07-05 & 48808.5 & MERLIN & & 4.993 & 15 & 175 &$<$230 & 46 & $\alpha$\\  
         1999-02-03 & 51212.5 & MERLIN & & 4.866 & 15 & 175 &$<$165 & 33 & $\beta$\\  
         2002-04-15 & 52379 & MERLIN & & 4.994 & 15 & 175 &$<$85 & 17 & $\gamma$\\
         2009-10-18 & 55122 & MERLIN & & 4.994& 16 & 286.5 & $<$99 & 33 & $\delta$\\
         2015-05-24 & 57166.042 & \textit{e}-MERLIN & CY2204 & 1.51 & 512 & 15.4 & $<$110 & 22\\
         2015-05-29 & 57171.543 & \textit{e}-MERLIN & CY2204 & 6.20 & 512 & 24.6 & 122$\pm$11 & 12 \\
         2015-05-30 & 57172.928 & \textit{e}-MERLIN & CY2204 & 4.88 & 512 & 15.6 & 174$\pm$15 & 9 \\
         2016-01-27  & 57414.239 & \textit{e}-MERLIN & CY3210 & 1.51 & 512 & 34.2 & $<$100 & 20 \\
         2016-05-10 & 57518.163 & \textit{e}-MERLIN & CY3210 & 5.07 & 512 & 22.7 & $<$60 & 12 \\          
          2021-03-06 & 59279.4 & EVN+\textit{e}-MERLIN & EM148 & 4.99 & 128 & 11.7 & 53$\pm$10 & 8.7\\
          2021-04-19 & 59323.377 & \textit{e}-MERLIN & CY11212 & 5.07 & 512 & 55 & $<$80 & 16 & \\
        \hline
         2015-05-30 & 57172.068 & \textit{e}-MERLIN & CY2204 & 5.60 & 1024 & 40.2 & 151$\pm$10 & 7 & *\\
         \hline
    \end{tabular}
    \caption{Observing log of M82 datasets presented in this work including upper limits from the literature for radio observations with resolution equivalent to, or greater than \textit{e}-MERLIN and an r.m.s. sensitivity $\leq$50$\mu$Jy~beam$^{-1}$. The column headings from left-to-right are: (1) The central day of the observation rounded down to the nearest day from the central MJD; (2) The central date in MJD of the observation but note that some observations from the literature were obtained over several weeks-months; (3) The array used in obtaining the data; (4) The project code of the dataset, if one is known; (5) The central frequency of the observing band in GHz; (6) the band width of the observation in MHz; (7) the total integration time on source in hours; (8) the integrated flux obtained from fitting a point source at the position of 41.37+60.2 if detected, or, an upper limit based on either the local r.m.s. sensitivity or for archival works, the detection limit used in that work; (9) the local r.m.s. sensitivity obtained from an off-source region near to 41.37+60.2; (10) Comments on the dataset including references to literature values: `$\alpha$': \citep{Muxlow1994}, `$\beta$': \citep{McDonald2001}, `$\gamma$': \citep{Fenech2008}, `$\delta$': \citep{Gendre2013}, `*':5$-$6 GHz combined dataset.
    }
    \label{tab:Rad_sourc_fluxes}
\end{table*}

Observations with the \textit{e}-MERLIN array were obtained in 2015 May at 1.51, 4.88 and 6.20\,GHz (project code CY2204, PI: Muxlow). The pointing centre was the well-known radio source 41.95+57.5, which for many years in the last century was the brightest compact source located within the M82 nuclear starburst \citep[see][for a discussion about the nature of this object]{Muxlow41.95}. The 4.88 and 6.20\,GHz were combined and calibrated separately to the 1.51\,GHz data. For simplicity, we refer to the combined 4.88 and 6.20\,GHz as the `5$-$6\,GHz 2015 \textit{e}-MERLIN dataset' here-after. We used the \textit{e}-MERLIN CASA Pipeline \citep[eMCP v1.1.09][]{eMCP} using \texttt{CASA} version 5.5.0 \citep{CASA} to calibrate the data and version 3.4 of the \texttt{wsclean} software \citep{offringa-wsclean-2014,offringa-wsclean-2017} to make large 1.33$\arcmin \times$1.33$\arcmin$ images of M82. The images were restored to a beam size of 50\,mas and 150\,mas in the 5$-$6\,GHz and 1.51\,GHz images, respectively. Multiple self-calibration loops were performed to improve the data quality, with the resulting images reaching $\leq$22$\mu$Jy~beam$^{-1}$ (see Table~\ref{tab:Rad_sourc_fluxes}) in each band. The 5$-$6\,GHz 2015 \textit{e}-MERLIN dataset produced an image with r.m.s. sensitivity 7$\mu$Jy~beam$^{-1}$ near to the source M82 X$-$1, a factor 2.5 better than previous observations with MERLIN \citep{Fenech2008}. In total, over 100 sources have been detected in either the 1.51 or the 5$-$6\,GHz 2015 \textit{e}-MERLIN dataset images, and these will be presented in a future work (Williams-Baldwin et al., in prep.). Additional data from 2016 (project code CY3210, PI: Muxlow) and 2021 (CY11212, PI: Williams-Baldwin) were reduced and self-calibrated in a similar way to the 2015 dataset, but due to a combination of shorter exposures, the lack of Lovell inclusion, or antenna failures, the resulting r.m.s. sensitivities are poorer (see Table~\ref{tab:Rad_sourc_fluxes}).

EVN+\textit{e}-MERLIN observations were obtained on 2021 March 6 under proposal code EM148 (PI: Muxlow). This project was designed with wide-field imaging in mind, utilising 64 channels in each of the four 32\,MHz wide sub-bands with 1s integration times at 6cm using the western EVN+\textit{e}-MERLIN telescopes, including: Jodrell Bank (Jb1), Westerbork (Wb), Effelsberg (Ef), Medicina (Mc), Onsala (O8), Torun (Tr), and the rest of the \textit{e}-MERLIN array. The data were calibrated in \texttt{AIPS} \citep{AIPS} using standard data reduction methods and imaged using the \texttt{AIPS} task \texttt{IMAGR}. In total, the EM148 project ran for 17 hours with 11.7 hours of M82 on-source time, resulting in an r.m.s. image sensitivity of 8.7$\mu$Jy~beam$^{-1}$ near to M82 X$-$1. All of the detected sources in the EM148 dataset will be presented in a future publication (Muxlow et al., in prep.).

We also provide detection limits of previous works in Table~\ref{tab:Rad_sourc_fluxes} from datasets with angular resolutions equivalent to the \textit{e}-MERLIN data (e.g. $\sim$50$-$200\,mas) between 1-6\,GHz, and sensitivities $\leq$50$\mu$Jy~beam$^{-1}$, quoting the relevant papers detection thresholds. All these works use a 5$\sigma$ detection threshold \citep{Muxlow1994,McDonald2001,Fenech2008} except \citealt{Gendre2013} who prefer a 3$\sigma$ threshold. We choose a minimum 5$\sigma$ detection threshold for our newly presented datasets, in order to be consistent with the majority of the literature and to rule out any low-level or spurious detections.

\begin{figure}
    \centering
    \includegraphics[width=\columnwidth]{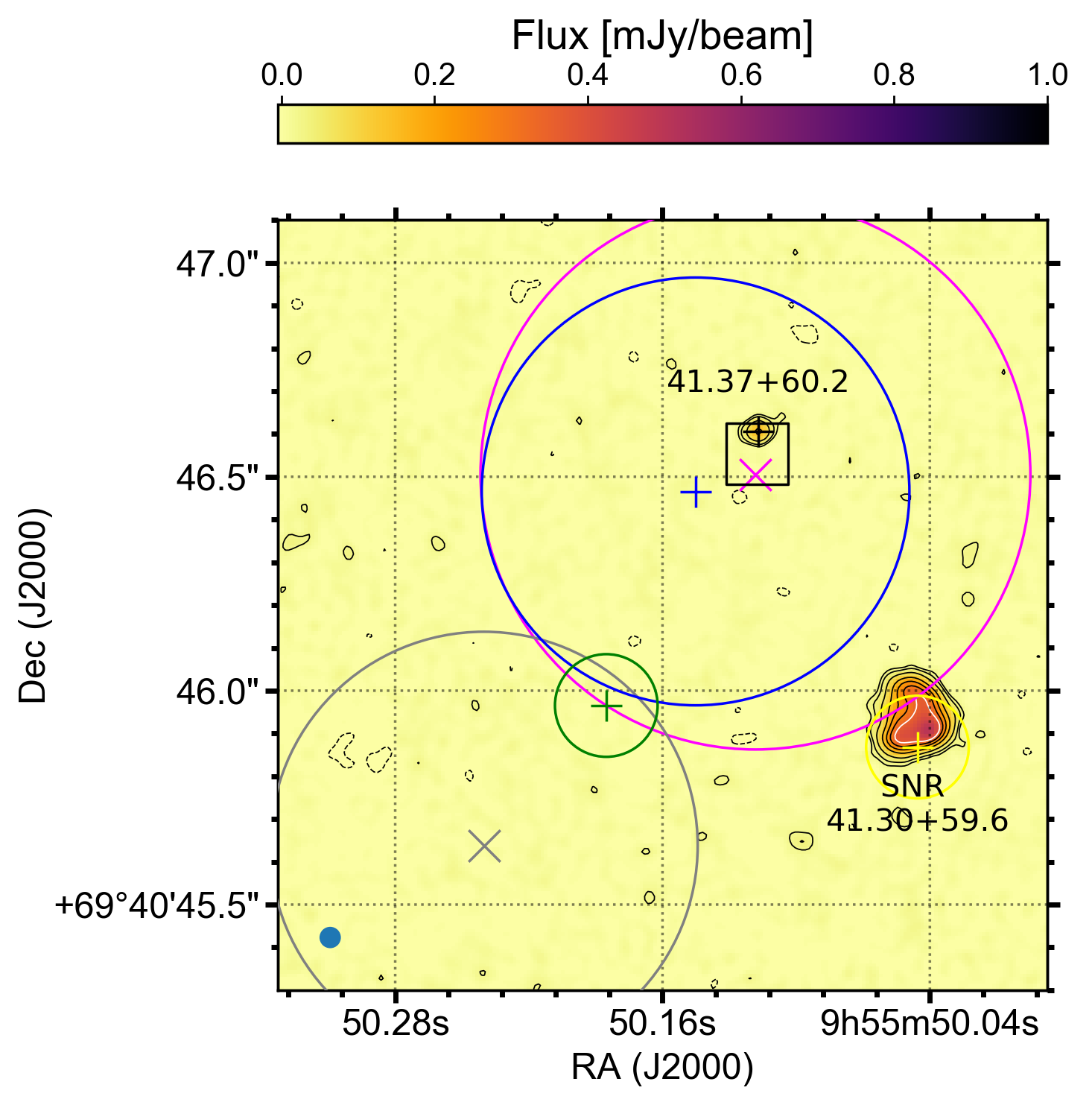}
    \caption{Image of the region surrounding the new source 41.37+60.2 (marked by a black `+' symbol) from the 5$-$6\,GHz 2015 \textit{e}-MERLIN dataset. The image size is 1.8$\arcsec\times$1.8$\arcsec$ and shows both 41.37+60.2 and the nearby SNR 41.30+59.6 to the south-west, labelled with a yellow plus-symbol and yellow circle showing the VLA synthesized beam size from \citealt{Kording2005}. The black contour levels are the image r.m.s. sensitivity 7$\mu$Jy~beam$^{-1} \times$ -3, 3, 5, 10, 20, 30, 40. A single white contour at 50$\times$7$\mu$Jy~beam$^{-1}$ is also shown for the SNR 41.30+59.6. The relevant positions and error circles noted in Section~\ref{sec:Xraypos} after astrometric corrections are shown in this image as follows: the magenta `X' and circle shows the X-ray position and positional uncertainty of M82 X$-$1, respectively, obtained from the sub-pixel method \citep{Xu2015}. The blue plus-symbol and circle shows the position and point spread function (psf) obtained from \citealt{Kording2005}. The X-ray source `S1' is denoted by the grey `X' and circle from \citealt{Xu2015}. The radio flare source (41.5+59.7) position and a 10$\%$ positional error from the Karl G. Jansky Very Large Array (VLA) 6~cm A configuration data is given by the green `+' and circle \citep{Kronberg1985}. The radio flare source has been attributed to the X-ray source S1 \citep{Xu2015}. The black square is the region of the 2021 EVN image shown in Figure~\ref{fig:Tom£VN}.}
    \label{fig:NS2localisation}
\end{figure}

\section{Results and Discussion}

Figure~\ref{fig:NS2localisation} shows our 5$-$6\,GHz 2015 \textit{e}-MERLIN dataset image near to the new radio source 41.37+60.2 which we label by the B1950 convention used for M82 sources \citep{Kronberg1985}. Figure~\ref{fig:NS2localisation} also shows the known SNR 41.30+59.6 to the south-west, the positions and error regions of known X-ray sources M82 X$-$1 and X-ray binary `S1' \citep{Xu2015}, plus the position of the radio transient 41.5+59.7 \citep[][see Figure caption and Section~\ref{sec:Xraypos} for details]{Kronberg1985}. 

\subsection{The new radio source 41.37+60.2: Source position and properties}

41.37+60.2 was initially found in the 5$-$6\,GHz 2015 \textit{e}-MERLIN dataset presented in this work. It resides at J2000 RA: 09$^{\rm h}$55$^{\rm m}$50$\fs$117, Dec: +69$^{\circ}$40$\arcmin$46$\farcs$60, to a positional accuracy of 5\,mas\footnote{The positional uncertainty is dominated by our ability to detect the source above the noise level. For a 5$\sigma$ source, the positional uncertainty is 10$\%$ \citep[see e.g., equation 1 in ][]{1998AJ....115.1693C}. As 41.37+60.2 has not been detected previously, we use this conservative 10$\%$ restoring beam positional uncertainty.}, lying approximately 0.8 arcsec north-east of the SNR 41.30+59.6 (see Figure~\ref{fig:NS2localisation}). Using the \texttt{CASA} task \texttt{IMFIT} and fixing the source size to the synthesized beam size (50\,mas), the source has an integrated flux \textit{S}$_{\rm \nu=5-6\,GHz}$ = 151$\pm$10$\mu$Jy corresponding to a $>$20$\sigma$ detection, including a 5$\%$ flux calibration error added in quadrature. The integrated fluxes in the two 2015 sub-bands are \textit{S}$_{\rm \nu=4.88\,GHz}$ = 174$\pm$15$\mu$Jy and \textit{S}$_{\rm \nu=6.20\,GHz}$ = 122$\pm$11$\mu$Jy. The noise levels in these maps are 9 and 12$\mu$Jy~beam$^{-1}$, respectively. We note that the 2015 integrated fluxes correspond to specific radio luminosities \textit{L}$_{\rm \nu=4.88\,GHz}$ = 1.0$\times$10$^{34}$ erg~s$^{-1}$ and \textit{L}$_{\rm \nu=6.20\,GHz}$ = 9.3$\times$10$^{33}$ erg~s$^{-1}$, for a distance of 3.2\,Mpc. Assuming \textit{S}$_{\rm \nu}\propto \nu^{\alpha}$, we fit a radio spectral index between these two bands resulting in a value of $-$1.48. This value is steep but consistent with optically-thin radio emission. If the source is variable within the observation, it may cause some artificial steepening of the radio spectrum. We searched for some intra-observation variability in the 5$-$6\,GHz 2015 \textit{e}-MERLIN dataset using the \texttt{-intervals-out} parameter in \texttt{wsclean}. This parameter enables `snap-shot imaging' to image the dataset into N intervals, where N is the number of intervals required. We used a N=5 to split the data into five approximately equal chunks to ensure good \textit{uv}-coverage and prevent significant reduction in sensitivity in each interval compared to the overall image sensitivity\footnote{See the \texttt{wsclean} documentation on usage of the \texttt{-intervals-out} parameter here: \url{https://wsclean.readthedocs.io/en/latest/snapshot_imaging.html}}. However we did not find any significant variation during the time period of either the 4.88\,GHz or 6.20\,GHz data. 

We searched for 41.37+60.2 in the 2015 1.51\,GHz dataset taken less than one week prior to the 5$-$6\,GHz 2015 \textit{e}-MERLIN dataset. Given the steep spectral index inferred above and assuming no source variability, we may expect to detect a mJy-level point source at the location of 41.37+60.2. The 1.51\,GHz data has an r.m.s. sensitivity of 22$\mu$Jy~beam$^{-1}$ and we do not detect it above a 5$\sigma$ detection threshold of 110$\mu$Jy, but a 3$\sigma$ contour encompasses the position of 41.37+60.2. Significant free-free absorption is known to affect many of the sources at sub-GHz frequencies in M82 \citep[e.g.][]{Wills408MHz,Varenius}. Using equations 2 and 3 of \citealt{Wills408MHz}, an emission measure of $>$1$\times$10$^7$cm$^{-6}$~pc is required to produce a non-detection at 1.51\,GHz for free-free absorption. This value is about an order of magnitude higher than sources with low frequency ($<$1\,GHz) detections in M82. However, most SNRs in M82 are not detected at sub-GHz frequencies, with implied emission measures of $>$10$^6$cm$^{-6}$~pc \citep{Wills408MHz,Varenius}. Therefore, while while it is possible that a spectral turnover below frequencies of 5\,GHz could reduce the flux of 41.37+60.2, source variability on short ($<$1 week) timescales is the most likely reason that 41.37+60.2 is not detected in the 1.51\,GHz 2015 data.

\begin{figure}
    \centering
    \includegraphics[width=\columnwidth]{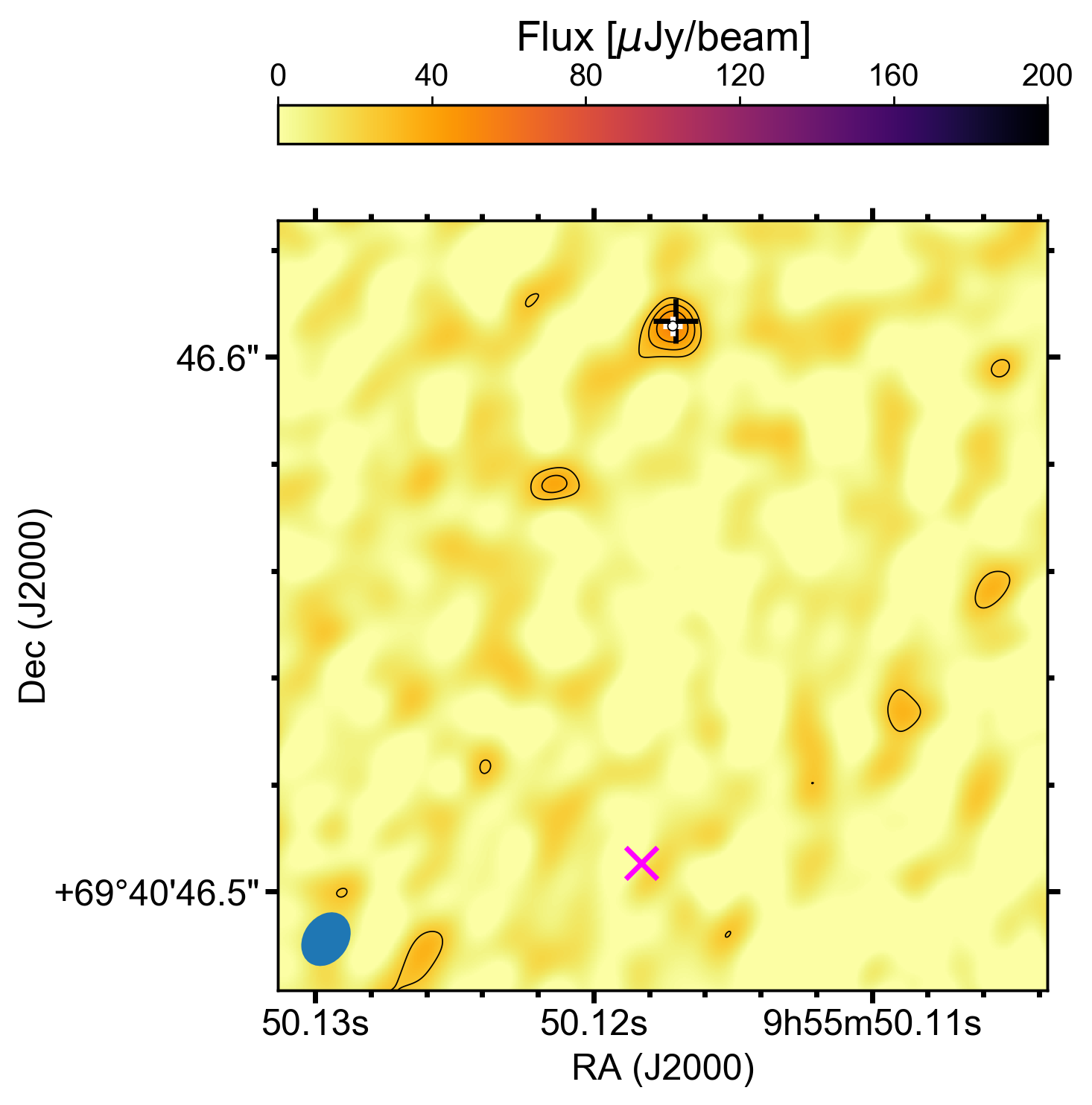}
    \caption{EVN+\textit{e}-MERLIN EM148 4.99\,GHz 2021 image of the square region in Fig~\ref{fig:NS2localisation} around the new radio source 41.37+60.2. The image size is 0.072$\arcsec\times$0.072$\arcsec$. The white plus-symbol shows the position of 41.37+60.2 from the EVN+\textit{e}-MERLIN 2021 data. The black plus-symbol shows the position of the 5$-$6\,GHz 2015 \textit{e}-MERLIN dataset once re-referenced to take into account the updated calibrator positions from the EVN+\textit{e}-MERLIN data. The size of the marks on both the EVN+\textit{e}-MERLIN and \textit{e}-MERLIN symbol represent the astrometric error of these datasets: $\pm$1.5\,mas and $\pm$5\,mas, respectively. The contour levels are at 8.7$\mu$Jy~beam$^{-1} \times$ -3, 3, 4, 5, 6. The magenta `X' symbol at the bottom of the image is the M82 X$-$ position from \citealt{Xu2015} after taking into account the astrometric corrections described in Section~\ref{sec:Xraypos}.}
    \label{fig:Tom£VN}
\end{figure}

The EVN+\textit{e}-MERLIN image (see Fig.~\ref{fig:Tom£VN}) obtained in 2021 has an r.m.s. sensitivity of 8.7$\mu$Jy~beam$^{-1}$ and a $>$6$\sigma$ point source co-incident with 41.37+60.2 is detected at a position of J2000 RA: 09$^{\rm h}$55$^{\rm m}$50$\fs$1172 ($\pm$0.0003\,sec), Dec: +69$^{\circ}$40$\arcmin$46$\farcs$605 ($\pm$1.5\,mas). The positional uncertainties are derived from the fitting errors using the \texttt{AIPS} task \texttt{JMFIT}, combined with a signal-to-noise error estimate of 10\% of the fitted beam, taken in quadrature, yielding an overall position error of $\pm$1.5\,mas in both Right Ascension and Declination. The integrated flux is \textit{S}$_{\rm \nu=4.99\,GHz, EVN}$ = 53$\pm$10$\mu$Jy, including a 5$\%$ flux calibration error added in quadrature. The source is unresolved with respect to the synthesized beam, which is 10.8$\times$8.4\,mas, position angle $-$35$^{\circ}$, limiting the physical size of the source to $<$0.16$\times$0.13\,pc, a specific radio luminosity of \textit{L}$_{\rm \nu=4.99\,GHz, EVN}$ = 3.2$\times$10$^{33}$ erg s$^{-1}$ and a brightness temperature of $T_{\rm B}$ $\geq$2.8$\times$10$^{4}$K. 

\begin{figure*}
    \centering
    \includegraphics[width=\textwidth]{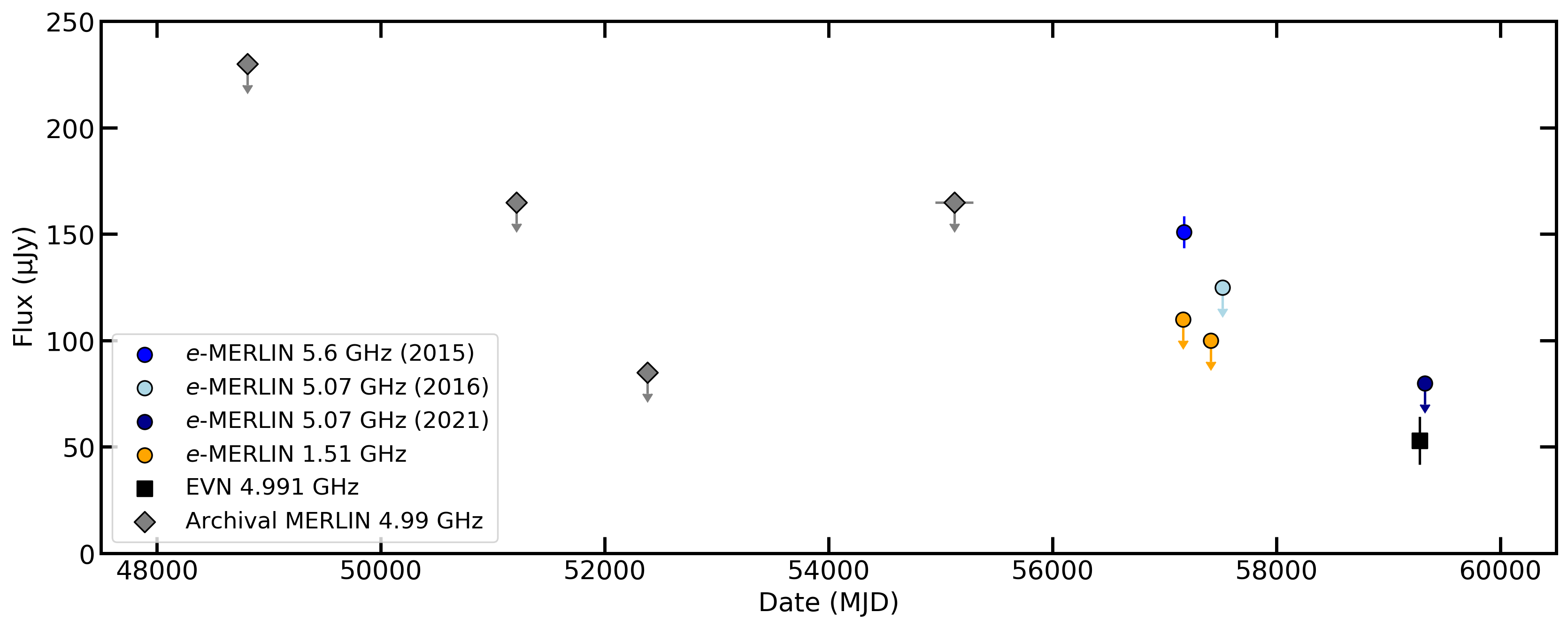}
    \caption{Inter-observation variability across the archival and new datasets presented in this work. Upper limits are denoted with downward facing arrows and the error bars in the x-axis refer to the total observing time of that observation. We only include the 5 GHz datasets in the archive from MERLIN observations in this plot as `Archival MERLIN 4.99 GHz' points for the following publications: \citealt[][See Table~\ref{tab:Rad_sourc_fluxes})]{Muxlow1994,McDonald2001,Fenech2008,Gendre2013}. For the 2015 \textit{e}-MERLIN 4.88 and 6.20\,GHz data, we provide a single data point, denoted `\textit{e}-MERLIN 5.6\,GHz (2015)' with a blue circle. }
    \label{fig:variable_LT}
\end{figure*}

41.37+60.2 does not appear in any high-resolution catalogues of M82 with MERLIN or the Karl G. Jansky Very Large Array \citep{McDonald,Fenech2008,Gendre2013}, despite these surveys having had the requisite sensitivity at $\sim$5\,GHz to have detected it. We show the detection limits of these observations in Table~\ref{tab:Rad_sourc_fluxes}. Comparing directly to the most sensitive archival MERLIN datasets at 5\,GHz, if the source was stable then it should have been detected at $\approx$11 and 6$\sigma$ levels in the \citealt{Fenech2008} and \citealt{Gendre2013} catalogues, respectively. However, we note that the aforementioned MERLIN images were combined from several observations and any radio variability may be averaged out from those datasets. 41.37+60.2 is not detected in the 2016 \textit{e}-MERLIN data which has 5$\sigma$ detection thresholds of 60 and 100$\rm \mu$Jy~beam$^{-1}$ at 5.07 and 1.51\,GHz respectively. It is also not detected in the 2021 \textit{e}-MERLIN dataset to a 5$\sigma$ detection threshold of 80$\rm \mu$Jy~beam$^{-1}$ at 5.07\,GHz. All of the above flux limits are shown in Fig.~\ref{fig:variable_LT}.

\subsection{The nature of the new radio source 41.37+60.2}

Having been detected in the 5$-$6\,GHz 2015 \textit{e}-MERLIN dataset and 2021 EVN+\textit{e}-MERLIN data, but not in archival or more recent datasets at similar frequencies and sensitivities, 41.37+60.2 is clearly a variable object over year-to-decade time-frames. Hence the source cannot be cataclysmic in nature. The EVN+\textit{e}-MERLIN data show that the source is compact in size ($<$0.16\,pc) and therefore is unlikely to be anything but a compact object. While the inferred brightness temperature ($T_{\rm B}$ $\geq$ 2.8$\times$10$^{4}$K) is at a similar level as a bright \ion{H}{II} region e.g., $T_{\rm B}$ $\sim$10000~K, the source variability argues against this interpretation. We stress that the brightness temperature derived here is a lower limit due to the source being unresolved. Moreover, the \ion{H}{II} regions in M82 have been observed to remain at a steady flux over several decades, so it is unlikely that this source is a \ion{H}{II} region. If the 1.51\,GHz 2015 observation is considered a non-detection, then the inferred rise-time ($<$1 week) and luminosity (\textit{L}$_{\rm \nu=4.88\,GHz}$ $\approx$1$\times$10$^{34}$ erg s$^{-1}$) of the radio emission are consistent with that of an X-ray binary or ULX origin if the radio observations were taken at the peak of a radio flare \citep{Pietka2015}. These radio luminosities are compatible with the radio luminosities observed in the brightest known X-ray binaries like Cyg X-3 \citep{Joseph2011}, which itself has been considered a `hidden' ULX previously \citep{Yang2023MNRAS.526L...1Y}. Assuming that the radio emission in 41.37+60.2 is due to an X-ray binary or ULX, then the compact and variable nature of the radio emission points towards an unresolved core, rather than part of an extended nebula as seen in some X-ray binaries \citep{Gallo2005Natur.436..819G,Motta2025,Atri2025} and ULXs \citep{Cseh2012, Berghea2020,Soria2020,Gong2023,Beuchert}. Indeed, the compact point source in the ULX Holmberg II X1 was shown to be variable over year-long time periods at VLBI resolutions with steep spectral indices that could only be explained by variability \citep{Cseh2014, Cseh2015}. Finally, the radio luminosities are consistent with other detected radio flaring ULX sources in the literature (see Section \ref{sec:nature}). Therefore, the radio emission in 41.37+60.2 is consistent with observed radio emission in X-ray binaries or ULXs.

\subsection{Astrometry and comparison to the X-ray position of M82 X$-$1}
\label{sec:Xraypos}

Between the epochs of the 5$-$6 GHz 2015 \textit{e}-MERLIN dataset in May 2015 and the 5\,GHz EVN+e-MERLIN (EM148) observations in March 2021, the phase-reference source common to both observation sets (J0955+6903) received an updated astrometric position of 09$^{\rm h}$55$^{\rm m}$33$\fs$173067 +69$^{\circ}$03$\arcmin$55$\farcs$06077 with a quoted error of $\pm$0.1 mas \citealt{Petrov2025}. The 5$-$6 GHz 2015 \textit{e}-MERLIN dataset positioning was upgraded to the new reference position, which resulted in a shift in RA of 2.2$\times$10$^{-6}$ seconds of Right Ascension and 
6.5\,mas in Declination. In the discussions below, we apply these relative offsets to all the source positions obtained from \textit{Chandra} and the Karl G. Jansky Very Large Array (VLA), to ensure a common reference frame between all datasets established from compact radio components within M82.

The radio position of 41.37+60.2 lies within an arcsecond of the literature positions of both the first radio transient found in M82, 41.5+59.7 \citep[][RA: 09$^{\rm h}$51$^{\rm m}$50$\fs$19, Dec: +69$^{\circ}$40$\arcmin$46$\farcs$0]{Kronberg1985,Kording2005} and the \textit{Chandra} X-ray position of M82 X$-$1 obtained from a sub-pixeling method \citep[][RA: 09$^{\rm h}$55$^{\rm m}$50$\fs$123, Dec: +69$^{\circ}$40$\arcmin$46$\farcs$54]{Xu2015} to within 70\,mas. \citealt{Kording2005} re-analysed the \citealt{Kronberg1985} data, as well as using new VLA data and \textit{Chandra} data, showing that the radio and X-ray positions between common compact sources were within good agreement to less than 0.1 arcsecond. The quoted uncertainty in the VLA and \textit{Chandra} positions of the compact sources is 0.1 arcsecond \citep{Kording2005} and this will dominate the astrometry when compared to higher resolution data, like our \textit{e}-MERLIN data. For example, the bright compact radio and X-ray source 41.95+57.5 differed in position by approximately 0.05 arcsec, much less than the 1 arcsecond point spread function (psf) of \textit{Chandra} and approximately one fifth of the beam size of the VLA. A full astrometric analysis aligning the \textit{Chandra} sources to the \textit{e}-MERLIN data is required to give a better positional accuracy, but this is not necessary for the analysis described below and is beyond the scope of this work.

Comparing the position of 41.95+57.5 between our \textit{e}-MERLIN 2015 data and that of \citealt{Kording2005}, we find a small difference of 0.005s in Right ascension and 0.04$\arcsec$ in Declination, equivalent to approximately 50\,mas, i.e., the \textit{e}-MERLIN synthesized beam in this work. We transformed the positions of the VLA data to the \textit{e}-MERLIN data and plot the positions of M82 X$-$1, 41.30+59.6 and the radio flare source 41.5+59.7 from \citealt{Kording2005} in Figure~\ref{fig:NS2localisation} as a blue, yellow and green plus-symbol, respectively. First, we note that the position of the SNR 41.30+59.6 (black plus-symbol) differs from the central shell in the \textit{e}-MERLIN image by $\sim$0.2 arcsec. This positional offset may be dominated by the aforementioned 0.1 arcsecond positional accuracy of the VLA and \textit{Chandra} data. Furthermore, \citealt{Fenech2008} noted that the source had undergone significant evolution in the preceding decade and the source is brighter in the south-west, which may explain the difference in position to the lower-resolution VLA data. Second, our new radio source 41.37+60.2 is 200\,mas away from the \citealt{Kording2005} M82 X$-$1 position and is the only source that resides within the 1 arcsecond psf region of M82 X$-$1. The radio flare source 41.5+59.7 is just outside the psf, 0.56 arcsec away from M82 X$-$1. 

We now concentrate on shifting the high-resolution \textit{Chandra} position of M82 X$-$1 obtained from a sub-pixeling method \citep{Xu2015} to the \textit{e}-MERLIN data. \citealt{Xu2015} also reported the discovery of an X-ray binary labelled `S1' approximately 1 arcsec south-east of M82 X$-$1, which they attributed to the radio flare source 41.5+59.7. We perform the same astrometric correction to the high-resolution \textit{Chandra} positions that were applied to the VLA/\textit{Chandra} data of \citealt{Kording2005}. This results in a positional offset between 41.37+60.2 and the \citealt{Xu2015} M82 X$-$1 position of 100\,mas, almost all of which is in declination. 
We plot these updated positions and 3$\sigma$ localisations of M82 X$-$1 and S1 from \citealt{Xu2015} in Figure~\ref{fig:NS2localisation} as magenta and grey `X' markers and circles, respectively. 
The position of 41.37+60.2 and the M82 X$-$1 position of \citealt{Xu2015} now disagree by 100\,mas, well within the 3$\sigma$ error circle for the \textit{Chandra} X-ray data.

To be certain that there is no chance alignment of radio and X-ray sources, we calculate the likelihood of detecting a radio source in M82 by chance within the error region of M82 X$-$1. First, we estimate the source density in M82 from our 5$-$6 GHz 2015 \textit{e}-MERLIN dataset. The compact sources in M82 span the inner 60$\arcsec$ $\times$ 15$\arcsec$, equal to 900$\arcsec ^2$ in area. There are 32 sources with detections above 20$\sigma$ like 41.37+60.2, equivalent to $\approx$0.036 per 1 $\arcsec ^2$. The X-ray error region of M82 X$-$1 is $\sim$0.5$\arcsec ^2$, so for any source within the 900$\arcsec ^2$ area, the probability of a radio source matching by chance is 2$\times$10$^{-5}$. If one considers the likelihood of detecting a radio source and the 10\,mas size of the EVN detection within the \textit{Chandra} region, this value falls to 6$\times$10$^{-9}$. Furthermore, if we follow the analysis of Section 3.1 in \citealt{Kording2005}, our \textit{e}-MERLIN data has a 50\,mas beam size, equating to 400 beams across the 1 arcsecond \textit{Chandra} psf. Assuming these beams are independent, the chance of a random 5$\sigma$ source in these beams is 0.02$\%$. As 41.37+60.2 is detected at $>$20$\sigma$ significance, the likelihood of this being random and not associated with M82 X$-$1 is vanishingly small. Therefore, we conclude that 41.37+60.2 is the radio counterpart of M82 X$-$1.

\subsection{X-ray variability of M82 X$-$1}
\label{sec:Brightmann}

The \textit{Chandra X-ray Observatory} and \textit{NuSTAR} telescopes have sufficient resolution to resolve the different X-ray sources in M82 providing high quality X-ray spectra. \citealt{Brightman2020} analysed quasi-simultaneous \textit{Chandra} and \textit{NuSTAR} data from 2015 January until 2016 October of M82 X$-$1 and M82 X$-$2, jointly-fitting the spectra between the two telescopes and providing X-ray flux measurements over the time period of our 2015 and 2016 \textit{e}-MERLIN data. 
In 2015 January, M82 X$-$1 had an X-ray flux of $S_{\rm X-ray~(0.5-30\,keV)}$ = 0.97$^{+0.04}_{-0.04}\times$10$^{-11}$~erg~cm$^{-2}$~s$^{-1}$, but in the following X-ray observation in 2015 June, this had increased by a factor four to $S_{\rm X-ray~(0.5-30\,keV)}$ =  4.26$^{+0.19}_{-0.13}\times$10$^{-11}$~erg~cm$^{-2}$~s$^{-1}$. The 2015 June data was obtained three weeks after the 5$-$6 GHz 2015 e-MERLIN dataset at a time when \textit{Swift/XRT} monitoring shows consistent X-ray flux around the level of $S_{\rm Swift~(0.5-8\,keV)} \sim$4$\times$10$^{-11}$~erg~cm$^{-2}$~s$^{-1}$ for several weeks beforehand \citep[see Figure 2 in][]{Brightman2020}. Converting the 4.26$^{+0.19}_{-0.13}\times$10$^{-11}$~erg~cm$^{-2}$~s$^{-1}$ flux into luminosity, assuming a distance, $d=$3.2\,Mpc results in a luminosity $L_{\rm X-ray~(0.5-30\,keV)}$ $\approx$ 5.2$\times$10$^{40}$~erg~s$^{-1}$.

In 2016, 41.37+60.2 was not detected to at 5$\sigma$ upper limit of 60$\mu$Jy. Comparing to the aforementioned \textit{Chandra, NuSTAR} and \textit{Swift/XRT} data from \citealt{Brightman2020}, we find that M82 X$-$1 was in a fainter state, with $S_{\rm Swift~(0.5-8\,keV)}\sim$2$\times$10$^{-11}$~erg~cm$^{-2}$~s$^{-1}$. The closest \textit{Chandra/NuSTAR} observations were obtained two weeks after the \textit{e}-MERLIN data on 2016 June 03 and provided a flux of $S_{\rm X-ray~(0.5-30\,keV)} = $ 2.31$^{+0.09}_{-0.06}\times$10$^{-11}$~erg~cm$^{-2}$~s$^{-1}$, equivalent to a luminosity of $L_{\rm X-ray~(0.5-30\,keV)}$ $\approx$ 2.8$\times$10$^{40}$~erg~s$^{-1}$.

The radio detection in 2015 coupled with the higher X-ray flux and the subsequent lack of radio detection in 2016 during a fainter X-ray state hints that the radio and X-ray emission may be coupled, with the radio emission only present in the X-ray bright states of M82 X$-$1. A dedicated monitoring programme is needed to test whether this tentative connection is real or not.

\subsection{The nature of M82 X$-$1: an intermediate-mass black hole, super Eddington neutron star or stellar-mass black hole?}
\label{sec:nature}
While the compact object at the heart of M82 X$-$1 is unknown, it is one of the best candidate IMBHs, with several authors suggesting a mass in the range 20--1000~M$_{\odot}$ \citep[see Figure~3 in][and references there-in]{Mondal}. The X-ray properties suggest a mass of $\sim$400~M$_{\odot} $\citep{Pasham2015,Motta2014}. 
But, as shown for the ULX 4XMM J111816.0$-$324910, using the X-ray variability and fluxes alone can lead to wildly different black hole mass measurements, depending on the (often strong) assumptions made \citep{Motta2020}. In the case of M82 X$-$1, we have shown that it has a radio counterpart (41.37+60.2). The combination of the compact nature of the radio emission, radio variability, and optically-thin spectral index of this radio component is consistent with radio emission observed in X-ray binaries or ULXs. We now compare the X-ray and radio data together to explore the possible nature of the compact object in M82 X$-$1.

We first compute the X-ray radio-loudness parameter \citep[$R_{\rm X}$=$\nu L_{\rm \nu (5~GHz)}/L_{\rm X-ray}$, see][]{Terashima}, which can be used as a proxy for radio-loudness when comparing different types of compact object. This metric was computed for four ULXs \citep{Mezcua2013MNRAS.436.1546M} using radio VLBI data, showing that in general they had values log($R_{\rm X}$) $\leq$ $-$4.4, with the IMBH candidate NGC 5457$-$X9 the exception, where a radio counterpart was detected with log($R_{\rm X}$) $\geq$ -4.1. Computing this value from the detection in 2015 of 41.37+60.2 yields log($R_{\rm X}$) = $-$5.9. However, this value is also consistent with X-ray binaries (log($R_{\rm X}$) $\leq$ $-$5.3), though not SNRs, young stars or a low-luminosity AGN \citep{Neff,Mezcua2013MNRAS.436.1546M}. 

In X-ray binaries, the radio:X-ray plane connects the disc and jet \citep[e.g.,][]{Corbel2000,FenderBelloni,Fender2004} in the form of a non-linear relationship between the `hard' X-ray spectral state and the radio emission emanating from a compact flat-spectrum jet \citep[e.g., ][]{GalloFenderPooley,Corbel2003,Coriat2011,Corbel2013}, commonly of the form $L_{\rm radio} \propto$~$L_{\rm X-ray}^{\sim 0.6}$, but some X-ray binaries show a steeper correlation $L_{\rm radio} \propto$~$L_{\rm X-ray}^{\sim 1.4}$ \citep[e.g., see ][]{Coriat2011}. ULXs follow their own spectral states \citep[e.g.,][]{Urquhart}, but if the compact object is a sub-Eddington IMBH, then it may fit along this correlation \citep{Panurach2024}. 

\begin{figure}
    \centering
    \includegraphics[width=\columnwidth]{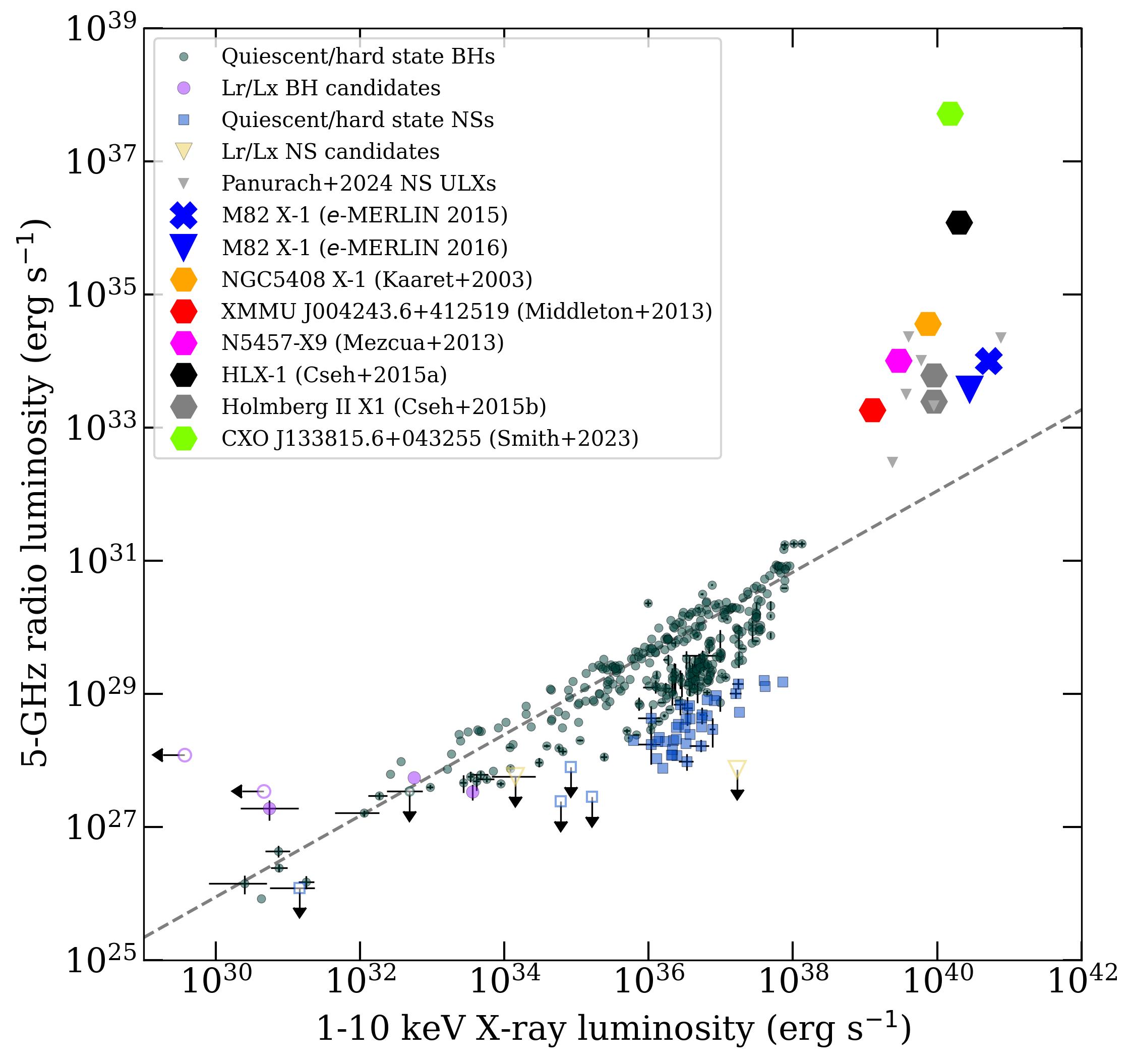}
    \caption{Radio:X-ray plane of X-ray binaries and other compact objects obtained from the online repository maintained by Arash Bahramian \citep{Bersavosh}, with the source types and symbols shown in the legend and the dark green dashed line representing the radio:X-ray correlation for black hole X-ray binaries of the form $L_{\rm radio} \propto$~$L_{\rm X-ray}^{0.61}$. We have included a sample of radio detections of intermediate-mass black hole candidates and radio upper limits of pulsating neutron star ULXs (dark grey downward facing triangles) from \citealt{Panurach2024}. Note that we have adjusted the radio flux for M82 X$-$2 to a 5$\sigma$ upper limit of 35$\rm \mu$Jy~beam$^{-1}$ using our deep \textit{e}-MERLIN 5$-$6\,GHz 2015 dataset presented here. Our \textit{e}-MERLIN radio detection of M82 X$-$1 is shown by a blue cross and the upper limit from the 2016 data is shown as a downward facing blue triangle. The radio detected intermediate-mass black hole candidates are described in Section~\ref{sec:nature}. Note that the radio and X-ray observations are obtained from sub-arcsecond resolution instruments, but may not be quasi-simultaneous with the X-ray measurements.}
    \label{fig:RX}
\end{figure}

In Figure~\ref{fig:RX}, we have added our radio detection in 2015 and upper limit from 2016 to the radio:X-ray plane for 41.37+60.2 including the X-ray luminosities calculated in Section~\ref{sec:Brightmann} from the \textit{Chandra/NuSTAR} data \citep{Brightman2020}. We include on this plot `hard'-state black hole and neutron star X-ray binaries. We also include the radio detected ULXs in the literature: NGC 5408 X$-$1 \citep[][]{Kaaret2003}, XMMU J004243.6+412519 \citep[][]{Middleton2013}, NGC 5457$-$X9 \citep[][]{Mezcua2013MNRAS.436.1546M}, Holmberg II X$-$1 \citep[][]{Cseh2015}, HLX$-$1 \citep[][]{2015CsehHLX1}, CXO J133815.6+043255 \citep[][]{Smith2023}, and the upper limits of pulsating ULXs obtained from \citealt{Panurach2024}. For M82 X$-$2, we use the upper limit on the radio flux from our 2015 dataset as it can be confused with a nearby \ion{H}{II} region and as noted by \citealt{Panurach2024}, this could lead to a significantly brighter radio flux than warranted by the observations. They also note that if a ULX is radio-detected then it is unlikely to be a neutron star, suggesting that the compact object is likely to be a black hole, either of stellar-mass or intermediate-mass origin. Our computed radio luminosities of the radio source 41.37+60.2 are consistent with all other radio-flaring ULX sources, all of which are suggested to be stellar or intermediate-mass black holes. 

Extending the radio:X-ray plane with a third term, the black hole mass, a `Fundamental Plane of Black Hole Activity' extends the radio:X-ray plane from stellar-mass black holes to super-massive black holes in galaxies \citep{Merloni,Falcke2004A&A...414..895F}. This `Fundamental Plane' assumes similar accretion and outflow mechanisms between these two types of objects, and whilst there are caveats to using it as a black hole mass estimator \citep[e.g. see][]{Gultekin2019}, it can be insightful to place an object onto this plane, given its radio and X-ray properties and assuming a black hole mass. 

The \citealt{Merloni} version of the plane converted to solve for the mass reads:
\begin{equation}
    \log (M_{\rm BH}) = 1.282 [\log (L_{\rm radio}) - 0.6 \log (L_{\rm X-ray}) - 7.33].
\end{equation}
where $M_{\rm BH}$ is the black hole mass in solar masses, $L_{\rm radio}$ is the radio luminosity in the 5\,GHz band and $L_{\rm X-ray}$ is the X-ray luminosity in the 2$-$10\,keV band. For the 2015 data, we use the 4.88\,GHz luminosity as a proxy for the 5\,GHz luminosity. For the X-ray luminosity, we use the WebPIMMS tool \citep{PIMMS} to estimate the fluxes in the 2$-$10\,keV from those derived using \textit{Chandra/NuSTAR} data in the 0.5$-$30\,keV band \citep{Brightman2020}, assuming a power law of 3, an average $n_{\rm H}$=1.3$\times$22~cm$^{-2}$. After performing this correction, the fluxes are appproximately five times smaller: 
$S_{\rm X-ray~(2-10\,keV)}$ $\sim$ 8.5$\times$10$^{-12}$~erg~cm$^{-2}$~s$^{-1}$ in 2015, corresponding to a luminosity of $L_{\rm X-ray~(2-10\,keV)} \sim $ 1$\times$10$^{40}$~erg~s$^{-1}$. We note that the spectral models fitted by \citealt{Brightman2020} are more complex than the simple absorbed power-law we have used, but we do not expect the fluxes to be significantly different from those we have calculated above. Substituting the above into equation (1), we arrive at a black hole mass of $M_{\rm BH} \sim$ 2650~M$_{\odot}$.

While suggestive of an IMBH, care must be taken in over-interpreting this black hole mass estimation. The nature of the compact object is unconfirmed and as the `Fundamental Plane' only applies for objects in the X-ray spectral `hard' state, the source could still be less massive accreting at the Eddington limit, or a larger mass object accreting at sub-Eddington rates \citep[see discussion in Section 3.8 of][]{Panurach2024}. Furthermore, our data are not simultaneous which may induce additional scatter \citep[see][for further discussion why simultaneity is required]{Gultekin2019}. Never-the-less, our radio detections of M82 X$-$1 are more likely to be from a BH, either stellar or intermediate-mass, than from a neutron star.

\section{Conclusions}

Using \textit{e}-MERLIN and EVN+\textit{e}-MERLIN data, we report a radio counterpart to the ultra-luminous X-ray source (ULX), M82 X$-$1. The radio source position (ICRF J2000 RA: 09$^{\rm h}$55$^{\rm m}$50$\fs$1172, Dec: +69$^{\circ}$40$\arcmin$46$\farcs$606 $\pm$1.5\,mas) lies within 100\,mas of the most accurate position of M82 X$-$1 in the literature \citep{Xu2015}. This source has an integrated flux $S_{\rm \nu=4.88\,GHz}$ = 174$\pm$15$\mu$Jy corresponding to a $>$20$\sigma$ detection in observations taken in 2015 May. Despite having requisite sensitivity, this source (41.37+60.2) was not catalogued by previous MERLIN observations taken in 2002 and 2009 \citep{Fenech2008,Gendre2013} and not detected in 2016 or 2021 \textit{e}-MERLIN data suggesting the source is transient or variable. A compact and unresolved radio source at the same position was detected with EVN+\textit{e}-MERLIN with integrated flux $S_{\rm \nu=4.99\,GHz}$ = 53$\pm$10$\mu$Jy, limiting the size of the source to $<$0.16$\times$0.13\,pc with a brightness temperature of $T_{\rm B}$ $\geq$2.8$\times$10$^{4}$K. 

We performed an astrometric analysis and found that our new radio source, 41.37+60.2, agrees to within 100 milliarcseconds with that of M82 X$-$1. We compared the radio variability and fluxes to other sources in the literature and found that the radio properties of 41.37+60.2 are similar to those of other radio-bright X-ray binaries and Ultra-luminous X-ray sources. By combining our radio data with X-ray data of M82 X$-$1 from the literature, we placed the source onto the radio:X-ray plane and `fundamental plane' of black hole activity. The data points reside in a similar position to other ULXs and stellar-mass black holes. Though care must be taken in obtaining a black hole mass using the `fundamental plane' alone, we arrive at a black hole mass of ${M}_{\rm BH} \sim$ 2650~M$_{\odot}$.

This work has highlighted the importance of regular monitoring of galaxies with a known history of a transient population with high-resolution high-sensitivity radio interferometers like \textit{e}-MERLIN and the EVN. 41.37+60.2 is the fifth radio transient/variable discovered in M82 that cannot be explained as an old supernova remnant or \ion{H}{II} region, following the X-ray binary 41.5+59.7 \citep{Kronberg1985}, the `MERLIN' transient \citep[43.78+59.3,][]{Muxlow2010}, the possible Gamma Ray Burst 41.95+57.5 \citep{Muxlow41.95} and the recent supernova SN2008iz \citep[][]{Brunthaler2009}. However, 41.37+60.2 is the first source in M82 associated with a ULX, and repeated monitoring of this source will help to understand the variability timescales of this source. Future radio telescopes like the SKAO and ngVLA will be crucial for detecting and monitoring more of these types of sources when they begin operations. 

\section*{Acknowledgements}

We would like to thank the anonymous reviewer for their helpful comments that have improved this work. DWB would like to thank Danielle Fenech, Matthew Middleton and Sara E. Motta for conversations that helped in the production of this manuscript. \textit{e}-MERLIN is a National Facility operated by the University of Manchester at Jodrell Bank Observatory on behalf of STFC, part of UK Research and Innovation. The European VLBI Network is a joint facility of independent European, African, Asian, and North American radio astronomy institutes. Scientific results from data presented in this publication are derived from the following EVN project code: EM148.

\section*{Data Availability}

The radio data can be made available on reasonable request to the corresponding author. Full radio maps of M82 will be made available in upcoming works (Williams-Baldwin et al., in prep. and Muxlow et al., in prep).



\bibliographystyle{mnras}
\bibliography{bib} 

\bsp	
\label{lastpage}
\end{document}